\documentstyle[12pt]{article}
\begin{document}

\baselineskip 0.6cm
\vspace{0.8cm}
\begin{center}
{\LARGE\bf LOOP VARIABLES FOR A CLASS
\vspace*{0.8cm}
  OF CONICAL SPACETIMES\footnote{ This work was supported in part by
Conselho Nacional de Desenvolvimento Cient\'{\i}fico e Tecnol\'ogico
(CNPq), Brazil.}}

\vspace*{1cm}

{\it V.B. Bezerra}\\
Blackett Laboratory, Imperial College\\
London SW7 \ 2BZ, U.K.\\

and\\

Departamento de F\'{\i}sica, CCEN\\
Universidade Federal da Para\'{\i}ba\\
58051-970 \ Jo\~ao Pessoa, Pb, Brazil

\vspace{0.8cm}

{\it P.S. Letelier}\footnote{e-mail: letelier@ime.unicamp.br}\\
Departamento de Matem\'atica Aplicada - IMECC\\
Universidade Estadual de Campinas\\
13081-970 \ Campinas, SP, Brazil
\end{center}
\vspace{0.5cm}

\begin{abstract}
Loop variables are used to describe the presence of topological
defects in spacetime. In particular we study the dependence of the
holonomy transformation on angular momentum and torsion for a
multi-chiral cone. We also compute the holonomies for multiple moving
crossed cosmic strings and two plane topological defects-crossed by
 a cosmic string.
\end{abstract}

\vfill

\hfill{June/95}

\newpage

\section{INTRODUCTION}
\renewcommand{\theequation}{1.\arabic{equation}}
\setcounter{equation}{0}

In the loop space formalism for gauge theories \cite{man1} the fields
 depend on paths rather than on spacetime points, and a gauge field is
described by associating with each path in spacetime an element of
the corresponding gauge groups. The fundamental quantity that arises
from this path-dependent approach, the non-integrable phase
factor \cite{wu} (or loop variable) represents the electromagnetic field or
a general gauge field more adequately than the field strength or the
integral of the vector potential \cite{wu}. In the electromagnetic case,
for example, as observed by Wu and Yang \cite{wu}, in a situation where
global aspects are taken into consideration the field strength
underdescribes the theory and the integral of the vector potential
for every loop overdescribes it. The exact description is given by
the factor $\exp \biggl(\frac{ie}{\hbar c} \oint_c A_\mu
 d x^\mu\biggr)$.

The extension of the loop formalism to the theory of gravity was
first considered by Mandelstam \cite{man2} who established several
 equations involving the loop variables, and also by Voronov and
Makeenko \cite{vor}.
Recently, Bollini et al. \cite{bol} computed the loop variables for the
gravitational field corresponding to the Kerr metric.

The loop variables in the theory of gravity are matrices representing
parallel transport along contours in a spacetime with a given affine
connection. They are connected with the holonomy transformations which
contain important topological information. These mathematical objects
contain information, for example, about how vectors change when parallel
transported around a closed curve. They also can be thought of as
measuring the failure of a single coordinate patch to extend the way
around a closed curve.

Suppose that we have a vector $v^\alpha$ at a point $P$ of a closed curve
$C$ in spacetime. Then, one can produce a vector $\bar{v}^{\alpha}$ at $P$
which, in general, will be different from $v^\alpha$, by parallel
transporting $v^\alpha$ around $C$. In this case, we associate with the
point $P$ and the curve $C$ a linear map $U^\alpha_\beta$ such that for
any vector $v^\alpha$ at $P$, the vector $\overline{v}^\alpha$ at $P$
 results from parallel transporting $v^\alpha$ around $C$ and is given by
$\overline{v}^\alpha = U^\alpha_\beta v^\beta$. The linear map
$U^\alpha_\beta$ is
called holonomy transformation associated with the point $P$ and the
curve $C$. If we choose a tetrad frame and a parameter $\lambda \in
[0,1]$ for the curve $C$ such that $C(0) =C(1) =P$, then in parallel
transporting a vector $v^\alpha$ from $C(\lambda)$
to $C(\lambda + d\lambda)$, the
vector components change by $\delta v^\alpha =
M^\alpha_\beta [x(\lambda)]v^\beta$,
where $M^\alpha_\beta$ is a linear map which depends on the tetrad, the
affine connection of the spacetime and the value of $\lambda$. Then, it
follows that the holonomy transformation $U^\alpha_\beta$ is given by the
ordered matrix product of the $N$ linear maps
\begin{equation}
U^\alpha_\beta = \lim_{N \rightarrow \infty} \prod^N_{i=1}
\biggl\{ \delta_{\alpha\beta} +
\frac{1}{N} M^\alpha_\beta [x(\lambda)]_{\lambda=i/N}\biggr\}
\end{equation}

One often writes the expression in Eq(1.1) as
\begin{equation}
U(C) = P\exp \biggl(\int_C M\biggr)
\end{equation}
where $P$ means ordered product along a curve C. Equation (1.2)
should be understood  as an abbreviation of the right hand side of
 Eq.(1.1). Note that if $M^\alpha_\beta$ is independent of $\lambda$,
then if follows from Eq.(1.1) that $U^\alpha_\beta$ is given by
$U^\alpha_\beta = (\exp M)^\alpha_\beta$.

In this paper we shall use the notation
\begin{equation}
U_{BA}(C) = P \exp\biggl( \int^B_A \Gamma_\mu(x(\lambda))
\frac{dx^\mu}{d\lambda} d\lambda\biggr)
\end{equation}
where $\Gamma_\mu$ is the tetradic connection and $A,B$ are the
initial and final points, respectively, of the path. Then, associated
with every path $C$ from a point $A$ to point $B$, we have a loop
variable given by Eq.(1.3) which is a function of the path $C$ as a
geometrical object.

The purpose of this paper is to examine the possibility of the use of
loop variables to characterize the presence of topological defects of
spacetime. In particular, we study a topological defect
corresponding to the multiple parallel chiral strings  in the
context of Einstein \cite{gale} and Einstein-Cartan
theories \cite{letor}, the multiple moving
crossed comic strings \cite{lega} in order to test if the strings are
really moving and crossing. Finally, we consider the topological defects
corresponding to two domain walls crossed by a cosmic string \cite{lewa}
 and two plane topological defects crossed by a cosmic string  \cite{lewa}
 also.

\section{LOOP VARIABLES IN THE SPACE-TIME OF MULTIPLE MOVING CROSSED
COSMIC STRINGS.}
\renewcommand{\theequation}{2.\arabic{equation}}
\setcounter{equation}{0}

Recently, Letelier and Gal'tsov \cite{lega} found an exact solution of the
Einstein equations describing an arbitrary number of non-parallel
straight infinitely long cosmic strings moving with different
constant velocities. The metric corresponding to this
configuration of strings is given by
 \begin{equation}
ds^2 = - e^{-4V} (dx + F_1dt + G_1dz)^2 - e^{-4V}(dy + F_2dt +
G_2dz)-dz^2+ dt^2
\end{equation}
where $V=2\sum^N_{i=1} \mu_i\ln r_i$, with $r_i=|\zeta
 -\alpha_i|$  and $\zeta=x+iy, \alpha_j=v_{xj}t +m_{xj} z+x_{0j}+
i( v_{yj}t +m_{yj}z+y_{0j})$. The functions $F_1 (G_1)$ and $F_2 (G_2)$
are the real and imaginary parts of two analytic functions on
the variable $\zeta$ also this functions depend on $t$ and $z$
through the combinations $\alpha_i$. The explicit form of this
functions will not play a major role in our analysis, they can
be found in \cite{lega}.

Our interest in compute the holonomies for this spacetime is to test
if really the strings are non-parallel and move in different
 directions. To do this  let us introduce a set of four vectors
 $e^\mu_{(a)} (a=1,2,3,4$ is a
tetradic index) which are orthonormal at each point with respect to
the metric with Minkowski signature, that is,
 $g_{\mu\nu} e^\mu_{(a)}e^\mu_{(b)}
= \eta_{ab} = \ {\rm diag} \ (-1,-1,-1,+1)$. We assume that the
$e^\mu_{(a)}$'s are matrix invertible, that is, that there exists an
inverse frame $e^{(a)}_\mu$ given by $e^{(a)}_\mu \ e^\nu_{(a)} =
\delta^\nu_\mu$ and $e^{(a)}_\mu \ e^\mu_{(b)} = \delta^a_b.$

Now define the 1-forms $\theta^a$ as
\begin{eqnarray}
\theta^1 &=& e^{-2V} (dx+ F_1dt + G_1dz)\nonumber\\
\theta^2 &=& e^{-2V} (dy+F_2dt +G_2dz)\nonumber\\
\theta^3 &=& dz \nonumber\\
\theta^4 &=& dt .
\end{eqnarray}
Then, in the coordinate system ($x^1=x, x^2=y, x^3=z,  x^4=t$), the
 tetrad frame defined by $\theta^a=e^{(a)}_\mu d x^\mu$
 is given by
\begin{eqnarray}
&&  e^{(1)}_1 = e^{-2V}, \ e^{(1)}_3 = e^{-2V}G_1, \; e^{(1)}_4
=e^{-2V}F_1\nonumber\\
&&  e^{(2)}_2 = e^{-2V}, \ e^{(2)}_3 = e^{-2V} G_2,\;  e^{(2)}_4
 =e^{-2V} F_2 \nonumber\\
&&  e^{(3)}_3 =1,\  e^{(4)}_4 =1.
\end{eqnarray}
Using Cartan's structure equations $d \theta^a
+\omega^a_{\,\,b}\wedge\theta^b =0$, for arbitrary
 functions $V, F_1, G_1, F_2$ and $G_2$ we get the
 following expressions for the tetradic connection
\begin{eqnarray}
&&\Gamma^1_{\mu 3} dx^\mu = - e^{-2V} \chi_1 dx +
\frac{1}{2} e^{-2V}\eta_1 dy - \frac{1}{2} e^{-2V}(\chi_2 - F_2\eta_1 +
2F_1\chi_1)dt +\nonumber\\
&& \hspace{2cm}\frac{1}{2} e^{-2V} (G_2\eta_1 -2G_1 \chi_1)dz = -
\Gamma^3_{\mu 1} dx^\mu \nonumber\\
&& \Gamma^2_{\mu 1} dx^\mu = -2 \frac{\partial V}{\partial y} dx - 2
\frac{\partial V}{\partial x} dy + \frac{1}{2} \biggl(\xi_1 + 4F_2
\frac{\partial V}{\partial x} -
 4F_1\frac{\partial V}{\partial y}\biggr) dt +\nonumber\\
&& \hspace{2cm}\biggl(2G_2\frac{\partial V}{\partial x} -
2G_1 \frac{\partial
V}{\partial y} +\xi_2\biggr)dz = - \Gamma^1_{\mu 2} dx^\mu\nonumber\\
&& \Gamma^2_{\mu 3} dx^\mu = \frac{1}{2} e^{-2V} (F_1\eta_1 +
2F_2\eta_1)dt + \frac{1}{2} e^{-2V}\eta_1 dx + e^{-2V}(\chi_3 dy +
\nonumber\\
&&  \hspace{2cm}\frac{1}{2} e^{-2V} (G_1\eta_1+ 2G_2 \chi_3)dz = -
\Gamma^3_{\mu 2} dx^\mu\nonumber\\
&& \Gamma^3_{\mu 4} dx^\mu = \frac{1}{2} e^{-4V} \eta_2 (dy + F_2dt +
G_2dz) + \frac{1}{2} e^{-4V} \chi_2 (dx + F_1dt + G_1dz) =
\Gamma^4_{\mu 3} dx^\mu\nonumber\\
&& \Gamma^4_{\mu 1} dx^\mu = e^{-2V} \chi_4 dx + \frac{1}{2} e^{-2V}
\eta_3 dy + \frac{1}{2} e^{-2V} (F_2\eta_3 + 2F_1\chi_4)dt + \nonumber\\
&&  \hspace{2cm}\frac{1}{2} e^{-2V} (G_2\eta_3 + \chi_2
 + 2G_2\chi_4)dz =  \Gamma^1_{\mu 4} dx^\mu\nonumber\\
&& \Gamma^4_{\mu 2} dx^\mu = \frac{1}{2} e^{-2V}
(2F_2\eta_4 + F_1\eta_3)dt + \frac{1}{2} e^{-2V}\eta_3 dx +
e^{-2V} \eta_4 dy +\nonumber\\
&& \hspace{2cm} \frac{1}{2} e^{-2V} (2G_2\eta_4 + \eta_2 + G_1\eta_3)dz =
\Gamma^2_{\mu 4} dx^\mu,
 \end{eqnarray}
where
\begin{eqnarray}
\eta_1 &=& \frac{\partial G_1}{\partial y} + \frac{\partial
G_2}{\partial x}\nonumber\\
\eta_2 &=& -\frac{\partial G_2}{\partial t} + \frac{\partial
G_2}{\partial x}F_1 + \frac{\partial G_2}{\partial y} F_2 -
\frac{\partial F_2}{\partial x}G_1 - \frac{\partial F_2}{\partial
y}G_2 +\frac{\partial F_2}{\partial z} \nonumber\\
\eta_3 &=& \frac{\partial F_1}{\partial y} + \frac{\partial
F_2}{\partial x}\nonumber\\
\eta_4 &=& 2\frac{\partial V}{\partial t} - 2\frac{\partial
V}{\partial x}F_1 - 2\frac{\partial V}{\partial y}F_2 +
\frac{\partial F_2}{\partial y} \nonumber\\
\chi_1 &=& 2\frac{\partial V}{\partial x}G_1 + 2\frac{\partial
V}{\partial y}G_2 - 2\frac{\partial V}{\partial z}- \frac{\partial
G_1}{\partial x} \nonumber\\
\chi_2 &=& -\frac{\partial G_1}{\partial t} + \frac{\partial
G_1}{\partial x}F_1 + \frac{\partial G_1}{\partial y} F_2 -
\frac{\partial F_1}{\partial x}G_1 - \frac{\partial F_1}{\partial
y}G_2 +\frac{\partial F_1}{\partial z} \nonumber\\
\chi_3 &=& 2\frac{\partial V}{\partial x}G_1 + 2\frac{\partial
V}{\partial y}G_2 - 2\frac{\partial V}{\partial z}- \frac{\partial
G_2}{\partial y} \nonumber\\
\chi_4 &=& 2\frac{\partial V}{\partial t} - 2\frac{\partial V}{\partial
x}F_1 - 2\frac{\partial V}{\partial y}F_2 + \frac{\partial
F_1}{\partial x} \nonumber\\
\xi_1 &=& \frac{\partial F_1}{\partial y} + \frac{\partial
F_2}{\partial x}\nonumber\\
\xi_2& = &\frac{\partial G_1}{\partial y} - \frac{\partial
 F_2}{\partial x}.
\end{eqnarray}
The property of analycity of the function $F(x+iy)=F_1+iF_2$ and
 similarly of  $G=G_1+iG_2$ tells us that
 \begin{equation}
\eta_1=\eta_3=0, \;\; \chi_1=\chi_3 .
\end{equation}
 We shall not use  explicitly this conditions in order to get
 general expressions for future reference.

Using the tetradic connections given by Eq.(2.4) we can compute the
loop variables. In our case we are interested in computing the loop
variables for segments in the $t$ and $z$ directions in order to
detect if the strings are moving with respect to each other and if
they are parallel or not. For a translation in time
$\Gamma_{\mu} dx^\mu = \Gamma_t dt$ with $\Gamma_t$ being
\begin{eqnarray}
\Gamma_t &=&\left(\begin{array}{cccc}
0 & B & C & A\\
B & 0 & -F & -D\\
C & F & 0 & E \\
A & D & -E & 0
\end{array}\right) \nonumber\\
\nonumber\\
\vspace*{0.3cm}
&=& - iFJ_{12}-iDJ_{13} - iEJ_{23}-iBJ_{41} - iCJ_{42}-iAJ_{43}  ,
\end{eqnarray}
where the boost parameters $A,B,C$ are given by
\begin{eqnarray}
&&A=\frac{1}{2} e^{-4V}(\eta_2 F_2 + \chi_2 F_1)\nonumber\\
&&B=\frac{1}{2} e^{-2V}(\eta_3 F_2 + 2 \chi_4 F_1)\nonumber\\
&&C=\frac{1}{2} e^{-2V}(2\eta_4 F_2 + \eta_3 F_1)
\end{eqnarray}
 and the rotation parameters $D,E,F$ are
\begin{eqnarray}
&&D=-\frac{1}{2}e^{-2V}(\chi_2 - \eta_1 F_2 + 2F_1\chi_1)\nonumber\\
&& E=\frac{1}{2}e^{-2V}(\eta_1 F_1 + 2\chi_3 F_2)\nonumber\\
&&F= -\frac{1}{2}
\biggl(\xi_1 + 4F_2 {\displaystyle\frac{\partial V}{\partial x}} - 4F_1
{\displaystyle\frac{\partial V}{\partial y}}\biggl) .
\end{eqnarray}

Then, for a segment in the time direction, the loop variable is a
combination of boosts in all directions and rotations. Note that
 the conditions (2.6) for the generic case do not eliminate any of
functions $A,...,F$.

Now, let us consider a segment in the $z$-direction. In this case we
have $\Gamma_{\mu 3} dx^\mu= \Gamma_z dz$, where
\begin{eqnarray}
\Gamma_z &=&\left(\begin{array}{cccc}
0 & B' & C' & A'\\
B' & 0 & -F' & -D'\\
C' & F' & 0 & E' \\
A' & D' & -E' & 0
\end{array}\right)\nonumber\\
\vspace*{0.3cm}
&=& - iF'J_{12}-iD'J_{13} - iE'J_{23}-iB'J_{41} - iC'J_{42}-iA'J_{43}
\end{eqnarray}
where
\begin{eqnarray}
A'&=& \frac{1}{2} e^{-4V}(\eta_2 G_2 + \chi_2 G_1), \;
B'= \frac{1}{2} e^{-2V}(\eta_3 G_2 + \chi_2+ 2\chi_4 G_1)\nonumber\\
C'&=& \frac{1}{2} e^{-2V}(2\chi_4 G_2 + \eta_2+\eta_3 G_1) , \;
D'= \frac{1}{2} e^{-2V}(\eta_1 G_2 - 2\chi_1 G_1)\nonumber \\
E' &=& \frac{1}{2} e^{-2V}(\eta_1 G_1 + 2\chi_3 G_2), \; F'= -2
\biggl(G_2\frac{\partial V}{\partial x} - G_1\frac{\partial V}{\partial
y}\biggr).
\end{eqnarray}
As in the previous case, the loop variable along the $z$ direction is
a combination of boosts and rotations which depends on the $z$
coordinate indicating that the cosmic strings crosses at some points,
and then they are not parallel. Similar result concerning the segment
in $t$-direction indicates that the strings are moving with respect
to each other.

\section{LOOP VARIABLES IN A MULTIPLE CHIRAL CONICAL SPACETIME.}

 \renewcommand{\theequation}{3.\arabic{equation}}
\setcounter{equation}{0}

In a recent paper Gal'tsov and Letelier \cite{gale} showed that the chiral
conical spacetime arises naturally from the spining particle
solution of (2+1)-dimensional gravity by an appropriate boost. This
chiral conical spacetime provides the gravitational counterpart for
the infinitely thin straight chiral strings in the same way that an
ordinary conical spacetime is associated with the usual string
\cite{gott} The
metric associated to the chiral conical spacetime \cite{gale} (spinning
string with cosmic dislocation) is given by
 \begin{equation}
ds^2 = - \overline{r}^{8\mu} (dr^2 + r^2 d\varphi^2)
 - (dz +4J^z d\varphi)^2 + (dt + 4J^t d\varphi)^2,
\end{equation}
where $J^t$ represents the string angular momentum, $2J^z/\pi$ is the
analogous of the Burgers-vector of dislocation and $\mu$ is the linear mass
density of the string. The angle $\varphi$ takes the values $0\leq
\varphi \leq 2\pi$, and the other variables: $-\infty < t< \infty,
 0 < r < \infty$, and
$-\infty < z < \infty$. If we consider a cartesian system of
coordinates $x=r\cos\varphi, y =r\sin\varphi$, we can write Eq.(3.1)
as
\begin{equation}
ds^2=- e^{-4V}(dx^2 +dy^2) - (dz + 4J^z\frac{ xdy-y dx}{r^2})^2
+(dt - 4J^t\frac{xdy-y dx }{r^2})^2
\end{equation}
with $V = 2\mu \ln r $.

The generalization of the chiral cone to a multiple chiral cone
 can \cite{gale}
be obtained by introducing the parameters $\mu_i, J^t_i, J^z_i,
i=1,2,\ldots,N$, defining each chiral string located at the points
$\vec{r}=\vec{r_i}$ of the plane $z=0$. The resulting metric has
the form of Eq.(3.2) with the following interchanges
\begin{eqnarray}
&&J^t\frac{ xdy-y dx}{r^2} \rightarrow \sum^N_{i=1}
 J^t_i \frac{(x-x_i)dy -(y-y_i)dx}{|\vec{r}-\vec{r_i}|^2}\nonumber\\
&&J^z \frac{ xdy-y dx}{r^2}\rightarrow \sum^N_{i=1} J^z_i
\frac{(x-x_i)dy -(y-y_i)dx}{|\vec{r}-\vec{r_i}|^2}\nonumber\\
 &&V= 2\mu \ln r \rightarrow V = \sum^N_{i=1} \mu_i
 \ln [r^2 -2rr_i\cos (\varphi -\varphi_i ) + r^2_i]
\end{eqnarray} 
As a consequence of Eq.(3.3), the spacetime generated by $N$ multiple
chiral cosmic string can be written as \cite{gale}
\begin{eqnarray}
&&ds^2 =  - e^{-4V}(dx^2 + dy^2) - [dz - \sum^N_{i=1} B_i(W^1_idy
- W^2_idx)]^2+\nonumber\\
&&\hspace{3cm}[dt -  \sum^N_{i=1} A_i(W^1_i dy - W^2_idx)]^2,
\end{eqnarray}
where
\begin{eqnarray}
&& A_i = 4J^t_i, \; \; B_i = 4J^z_i\nonumber\\
&& W^1_i = \frac{x-x_i}{|\vec{r}-\vec{r_i}|^2}, \;\;  W^2_i =
 \frac{y-y_i}{|\vec{r}-\vec{r_i}|^2}. \end{eqnarray}

In the previous case, satic one, the holonomy transformation was
calculated directly from the metric. For the present case, stationary one,
it is possible to do the same, but with a slightly redefinition of the
loop variables. This can be done because this solution of the Einstein
equation can be patched together from flat ccordinates patches but connected
by some additional matching condition in order to take into account the
helical structure and the shift in the $z$-direction.
First of all let us recover a previous result concerning the holonomy
in the static case \cite{bez}, specifically in the space of a multiple
 cosmic string
 \cite{letmu}. In this case one calculates the holonomy transformation
corresponding to circles in the $xy$-plane directly from the metric. Then,
when we parallel transport a vector around multiple cosmic strings at rest
at $\vec{r}=\vec{r_i}$ along a circle, this vector acquires a phase given by
 \cite{bez} $U(C)=\exp[-8\pi i (\sum_{j=1}^{N}\mu_j)J_{12}]$, where
$J_{12}$  is
the generator of rotations in the $xy$-plane, around the $z$-axis.
 Therefore, when
 we go around the multiple cosmic string from the point
$(\vec{x},t)$ to $(\vec{x'},t')$, the column vectors $(\vec{x},t)$
and $(\vec{x'},t')$ are related by
\begin{equation}
\left(\begin{array}{c}
x'\\
y'\\
z'\\
t'
\end{array}\right)=\left(\begin{array}{cccc}
\cos(8\pi\tilde{\mu})& \sin(8\pi\tilde{\mu})& 0&0\\
-\sin(8\pi\tilde{\mu} )& \cos(8\pi\tilde{\mu} )& 0&0\\
0&0&1&0\\
0&0&0&1\end{array}\right) \left(\begin{array}{c}
x\\
y\\
z\\
t
\end{array}\right),
\end{equation}
where we have set
\begin{equation}
\tilde{\mu}=\sum_{j=1}^{N}\mu_j.
\end{equation}

Since the space-time outside the multiple cosmic string is locally flat, we
 can describe the analytic solution purely in terms of spacetime patches with
Minkowski metric, but connected by some matching conditions which
 are given by Eq.(3.6), that relates points $(\vec{x},t)$ and $(\vec{x'},t')$
along the edges.

As in the multiple cosmic string case, the space-time of the multiple chiral
cosmic string is locally flat, and consequently we can describe it in
terms of space-time
patches with Minkowski metric, but connected by some conditions
 which are the same as
in the static multiple string case, except those concerning
the $t$ and $z$  coordinates.
These conditions are expressed by relating points  $(\vec{x},t)$
 and $(\vec{x'},t')$ as follows
\begin{eqnarray}
 x'&=&= \cos(8\pi\tilde{\mu})x+ \sin(8\pi\tilde{\mu})y\nonumber\\
y'&=&-\sin(8\pi\tilde{\mu})x+ \cos(8\pi\tilde{\mu})y\nonumber\\
z'&=&z +8\pi( \frac{r}{|\vec{r}-\vec{r_i}|^2})^2 J^z_i\nonumber\\
t'&=&t +8\pi( \frac{r}{|\vec{r}-\vec{r_i}|^2})^2 J^t_i,
\end{eqnarray}
where we are considering as paths circles in the $xy$-plane.

The transformations given by Eq.(3.8) can cast in the
 form of a homogeneous matrix multiplication as follows: let $M^B_A$ be
a five dimensional matrix, with $A$ and $B$ runing from 1 to 5. We take
$M^\mu_\nu$ equal to the rotation matrix given by \cite{bez},
 $U(C)=\exp(-8\pi
 i \tilde{\mu}J_{12}), \;\;M^4_5=8\pi( \frac{r}{|\vec{r}-
\vec{r_i}|^2})^2 J^t_i$
and $M^5_5=8\pi( \frac{r}{|\vec{r}-\vec{r_i}|^2})^2 J^z_i$, so that
\begin{eqnarray}
\left(\begin{array}{c}
x'\\
y'\\
z'\\
t'\\
1
\end{array}\right)&=&\left(\begin{array}{ccccc}
\cos(8\pi\tilde{\mu})& \sin(8\pi\tilde{\mu})& 0&0&0\\
-\sin(8\pi\tilde{\mu})& \cos(8\pi\tilde{\mu})& 0&0&0\\
0&0&1&0& 8\pi( \frac{r}{|\vec{r}-\vec{r_i}|^2})^2 J^z_i  \\
0&0&0&1&8\pi( \frac{r}{|\vec{r}-\vec{r_i}|^2})^2 J^z_i\\
0&0&0&0&1
\end{array}\right) \left(\begin{array}{c}
x\\
y\\
z\\
t\\
1
\end{array}\right)\nonumber\\
  \vspace*{0.5cm}
&=&\exp[ -8\pi( \frac{r}{|\vec{r}-\vec{r_k}|^2})^2 J^z_k M_3] \exp[
 -8\pi( \frac{r}{|\vec{r}-\vec{r_k}|^2})^2 J^t_k M_4] \times\nonumber\\
 & &\hspace{1cm}\exp(-8\pi\tilde{\mu}J_{12} )
\left(\begin{array}{c}
x\\
y\\
z\\
t\\
1
\end{array}\right),
\end{eqnarray}
where $ M_3 $ and $M_4$ are the following matrices
\begin{equation}
M_3= \left(\begin{array}{ccccc}
0&0&0&0&0\\
0&0&0&0&0\\
0&0&0&0&i\\
0&0&0&0&0\\
0&0&0&0&0
\end{array}\right), \;
M_4= \left(\begin{array}{ccccc}
0&0&0&0&0\\
0&0&0&0&0\\
0&0&0&0&0\\
0&0&0&0&i\\
0&0&0&0&0
\end{array}\right).
\end{equation}

Equation (3.9) is the exact expression for the holonomny for circles
 in the space-time. By defining $y^A=(y^\mu,1)$ we can cast the conditions
 (3.8) as $y'^A=M^A_By^B$ which tells us that the points
 $(\vec{x},t)$ and $(\vec{x'},t')$ along the edges
are related by the phase given by
Eq. (3.9) that depends on the parameters $\mu_j, J^t_j$ and $J^z_j .$

The existence of locally flat coordinates in this space-time
 permits us to consider Eq. (3.9) as a ``parallel transport" matrix. Then
we can say that when we carry a vector along a circle in this space-time it
 adquires a phase that depends on  $\mu_j, J^t_j$ and $J^z_j $
which prevents it to
be equal to the unit matrix. This effect is exclusively due to the
non trivial
topology of the space-time under consideration. This is a gravitational
analogue \cite{bez,bez2} of the Aharanov-Bohm effect \cite{ab}, but
in this case, purely at the classical level.

We can also compute the holonomy transformations for circles in
 the multiple
chiral conical space-time in the context of the Einstein-Cartan theory.
In this case the connection 1-forms appropriately chosen give
us \cite{letor}
\begin{equation}
\Gamma^1_{\mu 2} dx^\mu = 2(\partial_x V dy - \partial_y V dx) =
- \Gamma^2_{\mu 1} dx^\mu
\end{equation}
which can be written in cylindrical coordinates ($ x^1=r,
x^2=\varphi, x^3=z, x^4=t)$ as
\begin{equation}
\Gamma^1_{\mu 2} dx^\mu = -\frac{2}{r} \ \frac{\partial V}{\partial
\varphi} dr - \biggl(1 - 2r \frac{\partial V}{\partial r}\biggr) d\varphi =
- \Gamma^2_{\mu 1} dx^\mu.
\end{equation}
 Now, consider the same previous circle, at constant time. In  this
case $U_{2\pi,0}(C)$ is given by
\begin{equation}
U_{2\pi,0} (C) = \exp \biggl(\int^{2\pi}_0 \Gamma_{\varphi}
d\varphi\biggr) = \exp \biggl[-8\pi i \biggl(\sum^N_{j=1}
\mu_j\biggl) J_{12}\biggl]
\end{equation}
where
\begin{equation}
\Gamma_\varphi = i\biggl[1 - 4 \sum^N_{j=1} \mu_j
{\displaystyle\frac{R(R-r_i
\cos( \varphi-\varphi_i))}{(R^2 - 2R r_i \cos (\varphi-\varphi_i) +
r^2_i)}}\biggr] J_{12},
\end{equation}
R being the radius of the circle. Into Eq. (3.12)
we have dropped out the factor exp $(-2\pi
i J_{12})$ which is equal to the 4$\times$ 4 identity matrix. Note
that in this case the holonomy transformation has a simple
expression. It does not  carry information concerning angular momentum
and torsion and coincides with a previous result \cite{bez} concerning
the multiple cosmic string solution \cite{letmu}. Then, the concept of
holonomy can be used to detect different connections that come out
from Einstein and Einstein-Cartan theories.

\section{LOOP VARIABLES IN OTHER CONICAL SPACE-TIME.}
\renewcommand{\theequation}{4.\arabic{equation}}
\setcounter{equation}{0}

The purpose of this section is to complete a previous work \cite{bele}
in which we examen the possibility of the use of loop variables to
characterize the presence of topological defects of spacetime. We
considered a single domain wall crossed by multiple cosmic strings.
Other simple  examples of conical spacetimes include
two domain walls
crossed by a cosmic string of linear mass density $\mu$ and two
planes topological defects plus a cosmic string. A variety of
 conical space-time of diferent
topologies can be found in \cite{lewa}.

Now let us consider two domain walls parallels to the
$xy$-plane that intersect the $z$-axis at $\pm h$ and crossed by a single
cosmic string. The metric corresponding to the spacetime generated
by this configuration is

\begin{equation}
ds^2 = e^{-4\pi \sigma|h^2-z^2|}   - e^{4\pi\sigma t}
\rho^{-4\mu}(d\rho^2 +\rho^2d\varphi^2)-4z^2dz^2+dt^2 ],
\end{equation}
where $\sigma$ is the matter density of the wall and $\mu$ is the
linear mass density of the cosmic string.

Proceeding in the same way of the previous cases let us define the
appropriate 1-forms $\theta^a$ that give the usual flat spacetime limit
$\sigma=0, \mu=0$ as
\begin{eqnarray}
\theta^1 &=& e^{-2\pi \sigma|h^2-z^2|+2\pi\sigma t} (\rho^{-
2\mu}\cos\varphi d\rho -\rho^{-2\mu+1}\sin \varphi d\varphi)
\nonumber\\
\theta^2 &=& e^{-2\pi \sigma|h^2-z^2|+2\pi\sigma t}
(\rho^{-2\mu} \sin \varphi d\rho + \rho^{-2\mu+1}
\cos \varphi d\varphi)\nonumber\\
\theta^3 &=& 2z e^{-2\pi \sigma|h^2-z^2|}dz \nonumber\\
\theta^4 &=& e^{-2\pi \sigma|h^2-z^2|} dt
\end{eqnarray}

In a coordinate system $(\ x^1=\rho, \ x^2=\varphi, \ x^3=z, x^4=t )$,
the tetrad vectors are given by
\begin{eqnarray}
e^{(1)}_1& = &\rho^{-2\mu} e^{-2\pi\sigma|h^2-z^2|+2\pi \sigma t}
\cos \varphi ,\;\;
e^{(1)}_2 = -\rho^{-2\mu+1}  e^{-2\pi \sigma|h^2-z^2|+2\pi \sigma t}
\sin \varphi \nonumber\\
e^{(2)}_1 &=& \rho^{-2\mu} e^{-2\pi\sigma|h^2-z^2|+2\pi \sigma t}
\sin \varphi ,\;\;
e^{(2)}_2 = \rho^{-2\mu+1} e^{-2\pi\sigma|h^2-z^2|+2\pi \sigma t}
\cos \varphi \nonumber\\
 e^{(3)}_3 &= &2z e^{-2\pi \sigma|h^2-z^2|},\;\;
e^{(4)}_4 = e^{-2\pi \sigma|h^2-z^2|} .
\end{eqnarray}
Using these results we can compute the tetradic connections which are
\begin{eqnarray}
\Gamma^1_{\mu 2} dx^\mu &=& (1-4\mu)d\varphi = -\Gamma^2_{\mu 1}
dx^\mu\nonumber\\
\Gamma^1_{\mu 3} dx^\mu &=& 2\pi \sigma \frac{|h^2-z^2|}{h^2-z^2}
e^{2\pi\sigma t} \rho^{-4\mu} d\varphi = - \Gamma^3_{\mu 1}
dx^\mu \nonumber\\
\Gamma^2_{\mu 3} dx^\mu &=& 2\pi \sigma \frac{|h^2-z^2|}{h^2-z^2}
e^{2\pi\sigma t} \rho^{-4\mu+1} d\varphi = - \Gamma^3_{\mu 2}
dx^\mu\nonumber\\
\Gamma^4_{\mu 1} dx^\mu &=& 2\pi \sigma e^{2\pi\sigma t} \rho^{-4\mu}
d\varphi = \Gamma^1_{\mu 4} dx^\mu\nonumber\\
\Gamma^4_{\mu 2} dx^\mu &=& 2\pi \sigma e^{2\pi\sigma t}
\rho^{-4\mu+1}d\varphi = \Gamma^2_{\mu 4} dx^\mu\nonumber\\
\Gamma^4_{\mu 3} dx^\mu &=& 2\pi \sigma \frac{|h^2-z^2|}{h^2-z^2} dt
= \Gamma^3_{\mu 4} dx^\mu .
\end{eqnarray}

Let us consider the path $C$ as a circle centered on the $z$ axis with
radius $R$ lying on a plane parallel to the $xy$ plane at a fixed
time. In this case we have that $\Gamma_{\mu} dx^\mu=
\Gamma_{\varphi} d\varphi$ where
\begin{equation}
\Gamma_\mu= i(1-4\mu) J_{12} - 2\pi i \sigma e^{2\pi \sigma t}
\rho^{-4\mu +1} \biggl(J_{24} + \frac{|h^2-z^2|}{h^2-z^2} J_{23}\biggr).
\end{equation}
{}From Eq.(4.5) we get
\begin{equation}
U_{2\pi,0}(C) = \exp \biggl[-8\pi i \mu J_{12} - 4\pi^2 i \sigma
e^{2\pi \sigma t}\rho^{-4\mu+1} \biggl(J_{24} +
\frac{|h^2-z^2|}{h^2-z^2} J_{23}\biggr)\biggr].
\end{equation}
Equation (4.6) is the exact expression for the holonomy
transformation for a circle with center at the cosmic string and that
is parallel to the domain walls.

The holonomy transformation associated to the circle $C$ that
correspond to the domain walls only  $(\mu=0)$ and to the cosmic string
 only $(\sigma =0)$ are given, respectively, by
\begin{equation}
U_{2\pi,0}(C) = \exp \biggl[-4\pi^2 i \sigma e^{2\pi \sigma t}
\rho \biggl(J_{24} - \frac{|h^2-z^2|}{h^2-z^2} J_{23}\biggr)\biggr]
\end{equation}
and
\begin{equation}
U_{2\pi,0}(C) = \exp (-8\pi i \mu J_{12}).
\end{equation}
{}From these results we see that the holonomy transformations detect
the topological defects in all cases. In particular, for the domain
walls plus a cosmic string, the value of $U_{2\pi,0}(C)$ depends on
the radius of the circle. Note that $U_{2\pi,0}(C)$ distinguishes.

The regions $z<-h, -h<z<h$ and $z> h$ in the cases of two domain
walls crossed by  a cosmic string and two domain walls only.

In the case of two planes topological defects crossed by a cosmic
string with equation of state $p=-\gamma \sigma (\gamma < 1)$, the
metric so given by
\begin{equation}
ds^2 =  - \rho^{-8\mu} e^H(d\rho^2 + \rho^2 dy^2)+e^F(-4z^2dz^2+dt^2 ),
\end{equation}
where
\begin{eqnarray}
e^H& =& (t-|h^2-z^2|)^{\frac{1}{2(1-\gamma)}}\nonumber\\
e^F& = &\frac{1}{8\pi\sigma(1-\gamma)}\frac{1}{t-|h^2-z^2|} \biggl[
\frac{t-|h^2-z^2|}{t+|h^2-z^2|}\biggr]^{\frac{1}{4(\gamma -1)}}
 .
\end{eqnarray}

The holonomy transformation for the same circle $C$ described above
is given by
\begin{equation}
U_{2\pi,0}(C) = \exp \biggl[ -8\pi i \mu J_{12}- \frac{\pi i
e^{(H-F)/2}}{4(1-\gamma)(t-|h^2-z^2|)} \rho^{-4\mu+1} \biggl(J_{24} -
\frac{|h^2-z^2|}{h^2-z^2} J_{23}\biggr).
\end{equation}
To get Eq(4.11), we have used the following relations between the
tetradic connections
\begin{eqnarray}
\Gamma^1_{\mu 2} dx^\mu &=& (1-4\mu) dy = - \Gamma^2_{\mu 1}
dx^\mu\nonumber\\
\Gamma^1_{\mu 3} dx^\mu &=& \frac{e^{-F/2}}{2z} \ \frac{\partial}{\partial
z} (e^{H/2}) \rho^{-4\mu} d\rho = - \Gamma^3_{\mu 1} dx^\mu\nonumber\\
\Gamma^2_{\mu 3} dx^\mu &=& \frac{e^{-F/2}}{2z} \ \frac{\partial}{\partial
z} (e^{H/2}) \rho^{-4\mu+1} dy = - \Gamma^3_{\mu 2} dx^\mu\nonumber\\
\Gamma^4_{\mu 1} dx^\mu &=& e^{-F/2} \frac{\partial}{\partial t} (e^{H/2})
\rho^{-4\mu} d\rho = \Gamma^1_{\mu 4} dx^\mu\nonumber\\
\Gamma^4_{\mu 2} dx^\mu &=& e^{-F/2} \frac{\partial}{\partial t} (e^{H/2})
\rho^{-4\mu+1} dy = \Gamma^2_{\mu 4} dx^\mu\nonumber\\
\Gamma^4_{\mu 3} dx^\mu &=& \frac{e^{-F/2}}{2z} \ \frac{\partial}{\partial
z} (e^{F/2})dt + 2z e^{-F/2} \frac{\partial}{\partial t} (e^{F/2})dz=
\Gamma^3_{\mu 4} dx^\mu.
\end{eqnarray}

As in the previous case corresponding to two domain walls plus a
string, the holonomy transformation detects the topological defects
and distinguishes the different regions $z < -h, -h<z < h$ and $z >
h$.

\section{ CONCLUDING REMARKS}

We have shown by explicit computation from the metric corresponding
to a multiple parallel chiral cosmic strings that the loop
variables are combinations of rotations around the three axis and
boosts with appropriate parameters that depend on the characteristics
of each chiral string defined by $\mu_i, J_i^t$ and $J_i^z$. The
holonomy transformation, in this spacetime, assumes a simple form in
the context of the Einstein-Cartan theory which recover an expression
for the case of multiple cosmic strings. The loop variables
associated with the multiple moving crossed cosmic strings are
also combinations of rotations and boosts  in the directions of the
three spatial axis, and permit us to conclude that these are not
parallel and are moving with respect to each other.

In the of two domain walls and two planes topological defects plus
cosmic string, the holonomy transformation distinguishes the
presence of strings and membranes and depends on whether the loop
encircles the strings and in which side of the planes topological
defects are located.\\

\noindent
{\bf Acknowledgment}

One of us (VBB) wish to thank Prof. T.W.B. Kibble for reading the
manuscript, for helpful comments as well as for the hospitality at
Imperial College, where this work was completed.\\

\end{document}